\documentclass[12pt]{article}
\usepackage{amsfonts}
\usepackage{amssymb}
\usepackage{amsbsy}
\usepackage{mathrsfs}
\usepackage{amsmath}
\usepackage{graphicx}    
\usepackage{rotating}    
\usepackage{color}
\usepackage{epsfig}

\input epsf.sty

\newlength{\TZ}
\newcommand{\BEQ}{\begin{equation}}     
\newcommand{\BEA}{\begin{eqnarray}}
\newcommand{\BD}{\begin{displaymath}}
\newcommand{\EEQ}{\end{equation}}       
\newcommand{\EEA}{\end{eqnarray}}
\newcommand{\ED}{\end{displaymath}}
\newcommand{\vep}{\varepsilon}          
\newcommand{\vro}{\varrho}              
\newcommand{\D}{{\rm d}}                
\newcommand{\II}{{\rm i}}               
 
\newcommand{\demi}{\frac{1}{2}}         
\newcommand{\wit}[1]{\widetilde{#1}}    
\newcommand{\wht}[1]{\widehat{#1}}      
    
\renewcommand{\vec}[1]{\boldsymbol{#1}} 

\newcommand{\R}{\mathbb{R}}
\newcommand{\C}{\mathbb{C}}
\newcommand{\Z}{\mathbb{Z}}

\catcode`\@=11
\def\numberbysection{\@addtoreset{equation}{section}
        \def\theequation{\thesection.\arabic{equation}}}

\begin{document}

\begin{titlepage}

\vskip 1.5 cm
\begin{center}
{\LARGE \bf Meta-conformal invariance and the\\[0.16truecm] boundedness of two-point correlation functions}
\end{center}

\vskip 2.0 cm
\centerline{{\bf Malte Henkel}$^{a,b}$ and {\bf Stoimen Stoimenov}$^c$}
\vskip 0.5 cm
\centerline{$^a$Rechnergest\"utzte Physik der Werkstoffe, Institut f\"ur Baustoffe (IfB), ETH Z\"urich,}
\centerline{Stefano-Franscini-Platz 3, CH - 8093 Z\"urich, Switzerland}
\centerline{$^b$ Groupe de Physique Statistique, D\'epartement de Physique de la Mati\`ere et des Mat\'eriaux,}
\centerline{Institut Jean Lamour (CNRS UMR 7198), Universit\'e de Lorraine Nancy,}
\centerline{B.P. 70239, F -- 54506 Vand{\oe}uvre l\`es Nancy Cedex, France\footnote{permanent address}}
\vspace{0.5cm}
\centerline{$^c$ Institute of Nuclear Research and Nuclear Energy, Bulgarian Academy of Sciences,}
\centerline{72 Tsarigradsko chaussee, Blvd., BG -- 1784 Sofia, Bulgaria}

\begin{abstract}
The covariant two-point functions, derived from Ward identities in direct space, can be affected by consistency problems 
and can become unbounded for large time- or space-separations. This difficulty arises for several extensions of dynamical scaling,
for example Schr\"odinger-invariance, conformal Galilei invariance or meta-conformal invariance, but not for standard ortho-conformal invariance.  
For meta-conformal invariance in $(1+1)$ dimensions, which acts as a dynamical symmetry of a simple advection equation, 
these difficulties can be cured by going over to a dual space and an 
extension of these dynamical symmetries through the construction of a new generator in the Cartan sub-algebra. 
This provides a canonical interpretation of meta-conformally covariant two-point functions as correlators. Galilei-conformal correlators can be obtained
from meta-conformal invariance through a simple contraction. 
In contrast, by an analogus construction, Schr\"odinger-covariant two-point functions are causal response functions.  
All these two-point functions are bounded at large separations, for sufficiently positive values of the scaling exponents. 
\end{abstract}
\vfill
\end{titlepage}

\setcounter{footnote}{0}

\noindent
{\bf 1.} Dynamical symmetries are playing an useful r\^ole in the elucidation of the properties of many complex systems. 
One of the best-known examples is
{\it conformal invariance} in equilibrium phase transitions \cite{Bela84,diFran97}. When turning to dynamics, 
an often-studied case is the one of {\it Schr\"odinger-invariance} \cite{Henkel10}, 
which arises either from the free Schr\"odinger equation or else from the free diffusion equation, 
whose many applications to non-equilibrium dynamics include for example phase-ordering kinetics. 
One of the most elementary predictions of dynamical symmetries (such as conformal or Schr\"odinger-invariance) 
concerns the prediction of the form of the co-variant two-point functions, 
to be derived form the (e.g. conformal or Schr\"odinger) Ward identities \cite{Boyer67,Bela84}. 
These are built from quasi-primary scaling operators $\phi_i(t_i,\vec{r}_i)$, depending locally on a `time' coordinate 
$t_i\in\mathbb{R}$ and a `space' coordinate $\vec{r}_i\in\mathbb{R}^d$. 
Since both the conformal and the Schr\"odinger group contain time- and space-translations, and also spatial rotations, 
we can restrict to the difference $t:=t_1-t_2$ and the absolute value $r := |\vec{r}|=|\vec{r}_1-\vec{r}_2|$. 
Then, for conformal \cite{Poly70} and Schr\"odinger-invariance \cite{Henkel94}, 
respectively, the covariant two-point functions of {\em scalar} quasi-primary operators read, up to a global normalisation constant 
\begin{subequations} \label{1}
\begin{align}
\mathscr{C}_{12,\mbox{\footnotesize conf}}(t;\vec{r}) &= \left\langle \phi_1(t,\vec{r})\phi_2(0,\vec{0})\right\rangle =  
\delta_{x_1,x_2} \left[  t^2 + {r}^2 \right]^{- x_1}
\label{1C} \\
\mathscr{R}_{12,\mbox{\footnotesize Schr}}(t;\vec{r}) &= \left\langle \phi_1(t,\vec{r})\wit{\phi}_2(0,\vec{0})\right\rangle 
= \delta_{x_1,x_2} \delta({\cal M}_1+\wit{\cal M}_2)\,   t^{-x_1} \: \exp\left[ -\frac{{\cal M}_1}{2}\frac{{r}^2}{t}\right]
\label{1R}
\end{align}
\end{subequations}
A few more comments are required:
\begin{enumerate}
\item 
For conformal invariance, the properties of the two-point function, built from the scaling operators $\phi_i$, 
are described by the scaling dimensions $x_i$.
The two-point function is a {\it correlator} and is symmetric under permutation of 
the two scaling operators, viz. $\mathscr{C}_{12}(t;\vec{r})=\mathscr{C}_{21}(-t;-\vec{r})$. 
\item 
For Schr\"odinger-invariance, the two-point function is a (linear) {\it response function} 
-- it is cast here formally in the form of a correlator by appealing to Janssen-de Dominicis theory \cite{Taeu14} 
where one introduces a response operator $\wit{\phi}_i$ conjugate to the scaling operator $\phi_i$. 
The two-point function is now characterised by the pair $(x_i,{\cal M}_i)$ of a scaling
dimension and a mass ${\cal M}_i$ associated to each scaling operator $\phi_i$. For  `usual' scaling operators, masses are positive by convention, 
whereas response operators $\wit{\phi}_i$ have formally negative masses, viz. $\wit{\cal M}_i = - {\cal M}_i <0$. 
Because of causality, a response function $\mathscr{R}_{12}(t;\vec{r})$ is maximally asymmetric 
under permutation of the scaling operators, in the sense that for $t>0$, one has $\mathscr{R}_{12}\ne 0$, 
whereas as it must vanish, viz. $\mathscr{R}_{12}=0$, for $t<0$, because of causality. 
\end{enumerate}

In this Letter, we wish to point that a straightforwardish application of the Ward identities of time-space symmetries, often schematically reproduced
in the literature to derive results such as eqs.~(\ref{1}), implicitly assumes analyticity in the time-space arguments. 
If the two-point functions are not analytic, such an approach leads to 
inconsistencies.\footnote{An explicit example, leading to (\ref{meta2}), will be given below: a special case
is conformal galilean invariance.} 
As we shall show, in order to avoid this, a
certain extension of the dynamical symmetry is required and we shall indicate under which circumstances Ward identities can indeed be written down 
straightforwardly, such that the results are physically consistent. Such co-variant $n$-point functions arise for example
in several distinct non-relativistic versions of the AdS/CFT correspondence, 
see e.g. \cite{Barnich07a,Bagchi09,Bagchi09c,Fuertes09,Mart09,Duval09,Duval09b,Zhang10,Barnich11,Aizawa12,Minic12,Gray13,Barnich14,Henkel14b} 
and refs. therein.

\noindent {\bf 2.} Clearly, the result (\ref{1C}) of conformal invariance \cite{Poly70} is physically reasonable 
for a correlator and if $x_i>0$ it decays to zero for
large time- or space separations, viz. $|t|\to\infty$ or $r\to\infty$. On the other hand, 
the result (\ref{1R}) of Schr\"odinger-invariance \cite{Henkel94} does not explicitly contain the constraint $t>0$, 
physically required by causality. In addition, it is {\em not} obvious why the response
should vanish for large separations, even if $x_i>0$ is admitted. Of course, in the special case of Schr\"odinger-invariance, 
one might simply put in these features by hand. 
However, it is better to use the following, algebraically 
sound, procedure \cite{Henkel03a}: 
\begin{enumerate}
\item[(i)] consider the mass $\cal M$ as an additional coordinate. 
\item[(ii)] dualise by Fourier-transforming with respect to ${\cal M}_i$, 
which introduces dual coordinates $\zeta_i$. The terminology is borrowed from non-relativistic versions of the AdS/CFT correspondence.  
\item[(iii)] construct an extension\footnote{For the non-semi-simple Schr\"odinger algebra, 
the mathematical context are parabolic sub-algebras and their representations \cite{Knapp86}. 
Parabolic sub-algebras are made up from the Cartan sub-algebra $\mathfrak{h}$ and only `positive' generators. 
Our first construction of this kind still used the complete extension 
$\mathfrak{sch}(d)\subset\left(\mathfrak{conf}(d+2)\right)_{\mathbb{C}}$ into a conformal algebra in $d+2$ dimensions \cite{Henkel03a}. 
We realised later that the extension by a single generator $N$ is sufficient \cite{Henkel14a}, 
such that $\wit{\mathfrak{sch}}(d)$ is a non-trivial parabolic sub-algebra. If the extension by $N$ is not made, the physical consistency problems
remain \cite{Henkel14a}.} 
of the Schr\"odinger Lie algebra $\wit{\mathfrak{sch}}(d) := \mathfrak{sch}(d)\oplus \mathbb{C}N$, 
where the new generator $N$ is in the Cartan sub-algebra of $\wit{\mathfrak{sch}}(d)$. 
\item[(iv)] use the extended Schr\"odinger Ward identities, in the dual coordinates, to find the co-variant 
two-point function $\wht{\mathscr{R}}(\zeta_1-\zeta_2,t,\vec{r})$. 
\item[(v)] finally, transform back to the fixed masses ${\cal M}_i$. The result is \cite{Henkel03a,Henkel14a}
\BEQ \label{2R}
\mathscr{R}_{12,\wit{\mbox{\footnotesize Schr}}}(t;\vec{r}) = \left\langle \phi_1(t,\vec{r})\wit{\phi}_2(0,\vec{0})\right\rangle 
= \delta_{x_1,x_2} \delta({\cal M}_1+\wit{\cal M}_2)\,  \Theta({\cal M}_1 t)\,  t^{-x_1} \exp\left[ -\frac{{\cal M}_1}{2}\frac{{r}^2}{t}\right].
\EEQ
If one uses the convention ${\cal M}_1>0$, the Heaviside function $\Theta$ expresses the causality condition $t>0$. 
In addition, if $x_i>0$, the response function decays to zero for large
time- or space separations, as physically expected.

Comparing (\ref{2R}) with (\ref{1R}), we see the importance of formulating the extended Schr\"odinger Ward identities in dual space, 
where it is legitimate to treat $\wht{\mathscr{R}}$ as an {\it analytic function} of several variables. 
The extra generator $N$ provides an important restriction
such that in direct space, with the ${\cal M}_i$ fixed, it becomes explicit that $\mathscr{R}$ rather is a {\it distribution}. 

This comes about since the Ward identities lead to systems of first-order differential equations whose coefficients are holomorphic functions of
time and space. Hence its solution, the sought two-point function, will be holomorphic as well \cite{Hille97}. 
{\it A contrario}, if the physically required form of the
two-point function is not holomorphic, as in (\ref{2R}), a different form of the Ward identities must be sought. 
\end{enumerate} 

\noindent Here, we shall describe how to implement this procedure for a non-standard representation of 
conformal invariance which leads to a two-point correlation function distinct from (\ref{1C}). 
To be precise, we shall distinguish between ortho- and meta-conformal 
transformations.\footnote{From the greek prefixes $o\vro\theta o$: right, standard; and $\mu\vep\tau\alpha$: of secondary rank.}  

\noindent
{\bf Definition 1.} {\it (i) {\em Meta-conformal transformations}
are maps $(t,r)\mapsto (t',r')=\mathscr{M}(t,r)$, depending analytically on several parameters, such that they form a Lie group. 
The associated Lie algebra is isomorphic to the conformal Lie algebra $\mathfrak{conf}(d)$.  
A physical system is {\em meta-conformally invariant} if its $n$-point functions transform 
covariantly under meta-conformal transformations}. \\
{\it (ii) {\em Ortho-conformal transformations} (called `conformal transformations' for brevity) 
are those meta-conformal transformations $(t,r)\mapsto (t',r')=\mathscr{O}(t,r)$ 
which keep the angles in the time-space of points $(t,\vec{r})\in\mathbb{R}^{1+d}$ invariant.} 

In this Letter, we study the meta-conformal transformations, in $(1+1)$ time and space dimensions, 
with the following infinitesimal generators \cite{Henkel02,Henkel10}:
\BEA
X_n   &=& -t^{n+1}\partial_t-\mu^{-1}[(t+\mu r)^{n+1}-t^{n+1}]\partial_r -(n+1)\frac{\gamma}{\mu}[(t+\mu r)^{n}-t^{n}] -(n+1)xt^n  \nonumber\\
Y_{n} &=& -(t+\mu r)^{n+1}\partial_r- (n+1)\gamma (t+\mu r)^{n}
\label{infinivarconf}
\EEA
such that $\mu^{-1}$ can be interpreted as a velocity (`speed of light or sound') and where $x,\gamma$ are constants 
({\it `scaling dimension'} and {\it `rapidity'}\,). 
The generators obey the Lie algebra, for $n,m\in\mathbb{Z}$
\BEQ \label{commutators}
[X_n,X_{m}] = (n-m)X_{n+m},\quad  [X_n,Y_{m}] = (n-m)Y_{n+m},\quad [Y_n,Y_{m}] = \mu (n-m)Y_{n+m}
\EEQ
The isomorphism of (\ref{commutators}) with the conformal Lie algebra $\mathfrak{conf}(2)$ 
is seen as follows \cite{Henkel02,Henkel15}: 
write $X_n=\ell_n +\bar{\ell}_n$ and $Y_n=\mu \bar{\ell}_n$, 
where the generators $\langle \ell_n, \bar{\ell}_n\rangle_{n\in\Z}$ satisfy
$[\ell_{n},\ell_m]=(n-m)\ell_{n+m}$, $[\bar{\ell}_n,\bar{\ell}_m]=(n-m)\bar{\ell}_{n+m}$ and 
$[\ell_n,\bar{\ell}_m]=0$. If $\mu\ne 0$, the above Lie algebra (\ref{commutators})
is reproduced and hence is isomorphic to a pair of Virasoro algebras
$\mathfrak{vect}(S^1)\oplus\mathfrak{vect}(S^1)\cong\mathfrak{conf}(2)$ with a vanishing central charge.

The meta-conformal Lie algebra (\ref{commutators}) acts as a dynamical symmetry on the linear advection equation \cite{LeVeque99} 
\BEQ \label{ineq1}
{\cal S}\phi(t,r)=(-\mu\partial_t+\partial_r)\phi(t,r)=0 
\EEQ
in the sense that a solution $\phi$ of ${\cal S}\phi=0$, with scaling dimension  $x_{\phi}=x=\gamma/\mu$, 
is mapped onto another solution of the same equation. Hence 
the space of solutions of ${\cal S}\phi=0$ is meta-conformal invariant \cite{Henkel02} \mbox{(extended to Jeans-Poisson systems in \cite{Stoimenov15}). 
This follows from, with $n\in\mathbb{Z}$}  
\BEQ \label{dynsym}
{} [{\cal S},Y_n] = 0,\quad [{\cal S},X_{n}] = -(n+1)t^n{\hat S}+n(n+1)(\mu x-\gamma)t^{n-1}
\EEQ
Furthermore, the meta-conformal generators (\ref{infinivarconf}) make up one of only four closed Lie algebras 
of local time-space scaling transformations, constructed by requiring the
presence of time-translations with generator $X_{-1}$, dilatations with dynamical exponent $z$ and generator $X_0$ 
and spatial translations generated by $Y_{-1/z}$. 
The only three other closed Lie algebras which obey these requirements are 
ortho-conformal, conformal galilean and Schr\"odinger transformations \cite{Henkel02}. 

Now, quasi-primary scaling operators \cite{Bela84} are characterised by co-variance under the maximal finite-dimensional sub-algebra 
$\langle X_{\pm 1,0}, Y_{\pm 1,0}\rangle \cong \mathfrak{sl}(2,\R)\oplus\mathfrak{sl}(2,\R)$ for $\mu\ne 0$. Explicitly 
\BEA
X_{-1} &=& -\partial_t\;, \quad X_0 = -t\partial_t-r\partial_r-x\;, \quad
~X_{1} = -t^2\partial_t-2tr\partial_r-\mu r^2\partial_r-2xt-2\gamma r\nonumber\\
Y_{-1} &=& -\partial_r\;, \quad Y_0=-t\partial_r-\mu r\partial_r-\gamma\;, \quad
Y_1 = -t^2\partial_r-2\mu tr\partial_r-\mu^2r^2\partial_r-2\gamma t - 2\gamma\mu r ~~~~
\label{finitvarconf}
\EEA
Here, the generators $X_{-1}, Y_{-1}$ describe time- and space-translations, $Y_{0}$ 
is a (conformal) Galilei transformation, $X_{0}$ gives the
dynamical scaling $t\mapsto \lambda t$ of $r\mapsto \lambda r$  (with $\lambda\in\R$) such that the so-called `dynamical exponent' $z=1$
since both time and space are re-scaled in the same way and finally $X_{+1}, Y_{+1}$ give `special' meta-conformal transformations.
For a meta-conformally invariant system, quasi-primary operators $\phi_i$ are characterised by the 
parameters $(x_i,\gamma_i)$, while $\mu$ is simply a global dimensionful scale.
Using the generators (\ref{finitvarconf}) and extending to two-body generators $X_n^{[2]},Y_n^{[2]}$ in order to construct the 
meta-conformal Ward identities $X_n^{[2]} \left\langle \phi_1\phi_2\right\rangle=Y_n^{[2]}\left\langle \phi_1\phi_2\right\rangle=0$, one finds
the co-variant two-point function, up to normalisation \cite{Henkel02,Henkel10}
\BEQ \label{meta2}
\left\langle \phi_1(t_1,r_1) \phi_2(t_2,r_2) \right\rangle = 
\delta_{x_1,x_2}\delta_{\gamma_1,\gamma_2} \, (t_1-t_2)^{-2x_1} \left( 1 + \mu \frac{r_1-r_2}{t_1-t_2}\right)^{-2\gamma_1/\mu}
\EEQ
clearly {\em distinct} from the result (\ref{1C}) of ortho-conformal invariance. 
However, the result (\ref{meta2}) raises immediately the following questions:
\begin{enumerate}
\item is $\langle\phi_1\phi_2\rangle$ a correlator or rather a response, since neither of the symmetry or causality conditions are obeyed~?
\item even if $x_i>0$ and $\gamma_i/\mu>0$, why does $\langle\phi_1\phi_2\rangle$ 
not always decay to zero for large separations $|t_1-t_2|\to\infty$ or $|r_1-r_2|\to\infty$ ?
\item why is there a singularity at $\mu (r_1-r_2) = -(t_1-t_2)$ ?
\end{enumerate}
These difficulties, which cannot be fixed {\it ad hoc}, 
become even more pronounced in the `non-relativistic limit' $\mu\to 0$. From (\ref{infinivarconf},\ref{commutators}), 
one sees that the Lie algebra (\ref{commutators}) contracts to the non-semi-simple `altern-Virasoro algebra' $\mathfrak{altv}(1)$ 
(but without central charges). Its maximal finite-dimensional sub-algebra is
the {\it conformal galilean algebra}\footnote{\mbox{\sc cga}(d) is non-isomorphic to either the standard Galilei algebra or 
else the Schr\"odinger algebra $\mathfrak{sch}(d)$.} $\mathfrak{alt}(1)\cong\mbox{\sc cga}(1)\cong \mathfrak{bms}_3$ 
\cite{Havas78,Henkel97,Negro97,Henkel03a,Barnich07a,Bagchi09c}. 
The co-variant two-point function would become $\langle \phi_1\phi_2\rangle \sim \exp(-2\gamma_1 r/t)$ 
\cite{Henkel02,Bagchi09,Mart09} and does suffer from an analogous
difficulty with boundedness. However, for the {\sc cga}(d) algebra, it has been shown recently that a procedure analogous to
the one of the Schr\"odinger algebra, as outlined above, can be applied. This finally leads to
$\mathscr{C}=\langle \phi_1\phi_2\rangle \sim \exp\left(-2\left|\gamma_1 r/t\right|\right)$. Then boundedness is satisfied and 
$\mathscr{C}$ does obey the symmetry relations of a correlator \cite{Henkel15}. 

These two known examples might suggest that the correct identification of the Ward identities might only be possible for 
non-semi-simple Lie algebras, such as $\mathfrak{sch}(d)$ or $\mbox{\sc cga}(d)$. This is not true, as we shall now show. 
Our results will be summarised in Propositions 2 and 3.

\noindent {\bf 3.} Our construction follows the same steps as outlined above which have already been used to recast the co-variant two-point functions of 
Schr\"odinger- and conformal Galilean invariance into a physically reasonable form, see \cite{Henkel03a,Henkel14a,Henkel15}. 

First, we consider the `rapidity' $\gamma$ as a new variable. Second, it is dualised through a Fourier transformation, 
which gives the quasi-primary scaling operator
\BEQ
{\hat \phi}(\zeta, t, r)=\frac{1}{\sqrt{2\pi}}\int_{\R}\!\D\gamma\; e^{\II \gamma \zeta}\,\phi_{\gamma}(t,r)\label{Fouriergamma}
\EEQ
The representation (\ref{infinivarconf}) of the meta-conformal algebra becomes 
\BEA
X_n &=& \frac{\II(n+1)}{\mu}\left[\left(t+\mu r\right)^n - t^n\right]\partial_{\zeta} 
-t^{n+1}\partial_t -\frac{1}{\mu}\left[\left(t+\mu r\right)^{n+1}-t^{n+1}\right]\partial_r - (n+1)x t^n 
\nonumber \\
Y_n &=& \II(n+1)\left(t+\mu r\right)^n\partial_{\zeta} -\left(t+\mu r\right)^{n+1}\partial_r 
\label{dualvarconf}
\EEA
Third, we seek an extension of the Cartan sub-algebra $\mathfrak{h}$ by looking for a new generator $N$ such that
\BEQ \label{extensionCartan}
\left[X_n, N\right]=\alpha_n X_n \;\; , \;\; \left[Y_m, N\right]=\beta_m Y_m
\EEQ
where the constants $\alpha_n,\beta_m$ are to be determined. To find $N$, we start from the ansatz 
\BEQ 
N = A(\zeta,t,r,\mu)\partial_t+B(\zeta,t,r,\mu)\partial_r+C(\zeta,t,r,\mu)\partial_{\zeta}+D(\zeta,t,r,\mu)\partial_{\mu} + E(\zeta,t,r,\mu)
\label{findN}
\EEQ
which is slightly more general than the one used for the $\mbox{\sc cga}(1)$ \cite{Henkel15} 
and impose the commutators (\ref{extensionCartan}). The result is
\BD 
N=\alpha t\partial_t+\beta r\partial_r+(\beta-\alpha)\left((\zeta+c)\partial_{\zeta}-\mu\partial_{\mu}\right)+\nu,\label{Nfinal}
\ED
where $\alpha=\alpha_{-1}, \beta=\beta_{-1}, c, \nu$ are arbitrary constants. Since we seek an independent generator in $\mathfrak{h}$, we can
subtract the contribution proportional to $X_0$, retain as our definition of $N$
\BEQ \label{Ngenerator}
N := -r\partial_r -\left(\zeta+c\right)\partial_{\zeta} + \mu\partial_{\mu} - \nu
\EEQ
and then have, for $n\in\mathbb{Z}$
\BEQ \label{Ncommutators}
\left[ X_n, N \right] = 0 \;\; , \;\; \left[ Y_n, N \right] = - Y_n \;.
\EEQ
$N$ is a dynamical symmetry of (\ref{ineq1}), since $[{\cal S},N]=-{\cal S}$. 
This achieves the construction of the extended meta-conformal algebra $\wit{\mathfrak{conf}}(2) := \mathfrak{conf}(2) \oplus \mathbb{C} N$,
with commutators (\ref{commutators},\ref{Ncommutators}). \\

\noindent {\bf 4.} Co-variant two-point functions of quasi-primary scaling operators are found from the Ward identities
$X_n^{[2]}\left\langle\phi_1\phi_2\right\rangle=Y_n^{[2]}\left\langle\phi_1\phi_2\right\rangle=N^{[2]}\left\langle\phi_1\phi_2\right\rangle=0$,
with $X_n,Y_n,N\in\wit{\mathfrak{conf}}(2)$ and $n=\pm 1,0$ \cite{Henkel10}. Given the form of $N$, we also consider $\mu$ to a be further variable and set  
\BEQ \label{beginingtwo}
\langle{\hat \phi}(\zeta_1,t_1,r_1;\mu_1){\hat \phi}(\zeta_2,t_2,r_2;\mu_2)\rangle
={\hat F}(\zeta_1,\zeta_2,t_1,t_2,r_1,r_2;\mu_1,\mu_2)
\EEQ
Clearly, co-variance under $X_1$ and $Y_1$ implements time- and space-translation-invariance, such that
$\hat{F}=\hat{F}(\zeta_1,\zeta_2,t,r;\mu_1,\mu_2)$, with $t=t_1-t_2$ and $r=r_1-r_2$. Next, co-variance under $X_0$ and $Y_0$ produces
\BEA
\left(t\partial_t+r\partial_r+x_1+x_2\right){\hat F}(\zeta_1,\zeta_2,t,r;\mu_1,\mu_2) &=& 0 \label{scaling} \\
\left(t\partial_r+\mu_1 r\partial_r
-\II(\partial_{\zeta_1}+\partial_{\zeta_2})+(\mu_1-\mu_2)r_2\partial_r\right){\hat F}(\zeta_1,\zeta_2,t,r;\mu_1,\mu_2)&=&0
\label{covariancey0}
\EEA
Since ${\hat F}$ must not any longer depend explicitly on $r_2$, (\ref{covariancey0}) shows first that $\mu_1=\mu_2 =:\mu$ and 
\BEQ 
\left(t\partial_r+\mu r\partial_r-\II(\partial_{\zeta_1}+\partial_{\zeta_2})\right){\hat F}(\zeta_1,\zeta_2,t,r;\mu)=0.\label{covariancey02}
\EEQ
Similarly, co-variance under the special meta-conformal transformation $X_1,Y_1$ leads to 
\BEA
\left(\II r(\partial_{\zeta_1}-\partial_{\zeta_2})-t(x_1-x_2)\right){\hat F}(\zeta_1,\zeta_2,t,r;\mu)&=& 0   \label{covariancex1}\\
\left(t+\mu r\right)\left(\partial_{\zeta_1}-\partial_{\zeta_2}\right){\hat F}(\zeta_1,\zeta_2,t,r;\mu)&=&0. \label{covariancey1}
\EEA
Eq.~(\ref{covariancey1}) states that $(\partial_{\zeta_1}-\partial_{\zeta_2}){\hat F}=0$ such that $\hat{F}= {\hat F}(\zeta_+,t,r;\mu)$, with
$\zeta_+:=\demi(\zeta_1+\zeta_2)$. Then eq.~(\ref{covariancex1}) produces the constraint $x_1=x_2$. 

Finally, the required condition on the causality comes from co-variance under the $N$, which gives 
(we shall absorb from  now on $c$ into a translation of $\zeta_+$)
\BEQ \label{Ncovariance}
\left(r\partial_r+(\zeta_+ +c)\partial_{\zeta_+}-\mu\partial_{\mu}+\nu_1+\nu_2\right){\hat F}(\zeta_+,t,r;\mu)=0, 
\EEQ

\noindent {\bf 5.} The three conditions (\ref{scaling},\ref{covariancey02},\ref{Ncovariance}) fix the function
${\hat F}(\zeta_+,t,r,;\mu)$ which depends on three variables and the constant $\mu$, and also on the pairs of constants $(x_1,\nu_1)$ and
$(x_2,\nu_2)$ which characterise the two quasi-primary scaling operators $\hat{\phi}_{1,2}$. Solving eq.~(\ref{scaling}), it follows that
\BEQ \label{firstetap} 
{\hat F}(\zeta_{+},t,r;\mu) = t^{-2x}{\hat f}(u,\zeta_+,\mu)\;,\quad \mbox{\rm ~~ with $u=r/t$ and $x=x_1=x_2$.}   
\EEQ
Changing variables according to $v=\zeta_+ +\II u$ and $\hat{f}(u,\zeta_+,\mu)=\hat{g}(u,v,\mu)$, 
eqs.~(\ref{covariancey02},\ref{Ncovariance}) become 
\BEQ \label{Galileifinal}
\left(\frac{1+\mu u}{\mu u}\,\partial_u+\II \partial_v\right){\hat g}(u,v,\mu)=0 \;\; \mbox{\rm ~~and~~} \;\;
\left( u\partial_u+v\partial_v-\mu\partial_{\mu}+\nu_1+\nu_2\right){\hat g}(u,v,\mu)=0.
\EEQ
If $1+\mu u\ne 0$, the first equation (\ref{Galileifinal}) gives ${\hat g}(u,v,\mu)={\hat G}(w,\mu)$, where $w$ is obtained from
\BEQ \label{scalinvariable}
w=\int\!\D u\: \frac{\mu u}{1+\mu u} +\II \int \!\D v= u-\frac{\ln(1+\mu u)}{\mu}+\II v .
\EEQ
Inserted in the other eq.~(\ref{Galileifinal}), this leads to $(w\partial_{w}+\nu_1+\nu_2){\hat G}(w,\mu)=0$, hence 
\BEQ\label{resultfinal} 
{\hat G}(w,\mu)={\hat G}_0(\mu)w^{-\nu_1-\nu_2}
={\hat G}_1(\mu)\left(\zeta_+ +\II\frac{\ln(1+\mu u)}{\mu}\right)^{-\nu_1-\nu_2}.
\EEQ
Since $\mu$ is merely a parameter, ${\hat G}_1(\mu)$ is just a normalisation constant. We have proven  
 
\noindent{\bf Proposition 1}.{\em The dual two-point function, covariant under the generators $X_{\pm 1,0}, Y_{\pm 1,0}, N$ 
of the dual representation (\ref{dualvarconf},\ref{Ngenerator}) of the meta-conformal algebra 
$\wit{\mathfrak{conf}}(2)$, is up to normalisation} 
\BEQ \label{dualtwopoint}
{\hat F}(\zeta_1,\zeta_2,t,r)=\langle{\hat \phi}_1(t,r,\zeta_1){\hat\phi}_2(0,0,\zeta_2)\rangle
=\delta_{x_1,x_2}
\:|t|^{-2x_1}\left(\frac{\zeta_1+\zeta_2}{2} +\II\frac{\ln(1+\mu r/t)}{\mu}\right)^{-\nu_1-\nu_2}.
\EEQ
  
\noindent {\bf6.} To un-dualise, that is to carry out the the inverse Fourier transform on the dual two-point function (\ref{dualtwopoint}), 
a precise mathematical fact is required. 
We write $\mathbb{H}_+$ ($\mathbb{H}_{-}$) for the upper (lower) complex half-plane $w=u+\II v$ with $v>0$ ($v<0$).
Recall from \cite[ch. 11]{Akhiezer}:
  
\noindent{\bf Definition 2.} {\em A holomorphic function $g : \mathbb{H}_+\to\C $ is in the {\em Hardy class $H_2^+$}
if the bound 
\BEQ 
M^2:= \sup_{v>0}\int_{\R} \!\D u\: |g(u+\II v)|^2<\infty \label{carreintegrable}.
\EEQ
holds true. Analogously, a holomorphic function  $g : \mathbb{H}_-\to\C $ is in the {\em Hardy class $H_2^-$}, when the 
supremum in (\ref{carreintegrable}) is taken over $v<0$.}
  
\noindent{\bf Lemma 1}. \cite{Akhiezer} {\em If $g\in H_2^{\pm}$, then there are square-integrable functions 
${\cal G}_{\pm}\in L^2(\mathbb{R}_+)$, such that for $v>0$, one has the integral representation
\BEQ \label{integralrepresentation}
g(w)=g(u\pm\II v)=\frac{1}{\sqrt{2\pi}}\int_0^{\infty} \!\D\gamma\: e^{\pm\II \gamma w}{\cal G}_{\pm}(\gamma).
\EEQ}
  
To use this Lemma, we fix 
\BEQ \label{3.20}
\lambda:=\frac{\ln(1+\mu r/t)}{\mu}
\EEQ 
and recalling (\ref{dualtwopoint}), we write ${\hat F}=\delta_{x_1,x_2}\,|t|^{-2x_1}{\hat f}(\zeta_++\II \lambda)$ such that 
\BEQ  \label{f30}
f_{\lambda}(\zeta_+) := {\hat f}(\zeta_+ +\II \lambda) = \left( \zeta_+ +\II \lambda \right)^{-\nu_1-\nu_2} 
\EEQ 
\noindent {\bf Lemma 2.} \cite{Henkel15} {\it Let $2\nu := \nu_1+\nu_2>\demi$. 
Then $f_{\lambda}\in H_2^{+}$ for $\lambda>0$ and $f_{\lambda}\in H_2^{-}$ for $\lambda<0$.}

\noindent{\bf Proof:} Since $f_{\lambda}$ is holomorphic where it is defined, it remains to check the bound (\ref{carreintegrable}). Clearly, 
$\left|f_{\lambda}(u+\II{v})\right|=\left| (u+\II( v+\lambda))^{-2\nu}\right| = \left( u^2 + ({v}+\lambda)^2 \right)^{-\nu}$. 
For $\lambda>0$, by explicit computation    
\BD
M^2 = \sup_{{v}>0} \int_{\mathbb{R}} \!\D u\: \left| f_{\lambda}(u+\II {v})\right|^2 
=  \frac{\sqrt{\pi\,}\: \Gamma(2\nu-\demi)}{\Gamma(2\nu)} 
\sup_{{v}>0} \left({v}+\lambda\right)^{1-4\nu}
< \infty
\ED
since the integral converges for $\nu>\frac{1}{4}$. For $\lambda<0$, the argument is similar. \hfill ~ q.e.d.

Therefore, we have from Lemma 1, eq.~(\ref{integralrepresentation}), where $\Theta$ is the Heaviside function   
\BEQ \label{3.21}
\sqrt{2\pi}\,{\hat f}(\zeta_+ +\II\lambda)=
 \Theta(\lambda)\int_0^{\infty}  \!\D\gamma_+\: e^{+\II(\zeta_+ +\II\lambda)\gamma_+}{\hat {\cal F}}_{+}(\gamma_+)  
+\Theta(-\lambda)\int_0^{\infty} \!\D\gamma_+\: e^{-\II(\zeta_+ +\II\lambda)\gamma_+}{\hat {\cal F}}_{-}(\gamma_+).  
\EEQ

\noindent To carry out the back-transformation, we distinguish the cases $\lambda>0$ and $\lambda<0$. 
If one has $\lambda>0$, we have from (\ref{3.20},\ref{3.21}), with $\zeta_{\pm}=\demi(\zeta_1\pm\zeta_2)$
\BEA 
F &=& \frac{|t|^{-2x}}{\pi\sqrt{2\pi}}\, {\hat G}_1(\mu)\int_{\R^2} \!\D \zeta_+ \D\zeta_-\; 
e^{-\II(\gamma_1+\gamma_2)\zeta_+}\, e^{-\II(\gamma_1-\gamma_2)\zeta_-}
\int_{\R}\!\D\gamma_+\,\Theta(\gamma_+){\hat {\cal F}}_+(\gamma_+)e^{-\gamma_+\lambda}e^{\II\gamma_+\zeta_+}\nonumber\\
&=& \frac{|t|^{-2x}}{\pi\sqrt{2\pi}}\,{\hat G}_1(\mu)\int_{\R} \!\D\gamma_+\,\Theta(\gamma_+){\hat {\cal F}}(\gamma_+)e^{-\gamma_+\lambda}
\int_{\R} \!\D\zeta_{-}\; e^{-(\gamma_1-\gamma_2)\zeta_-}\int_{\R} \!\D\zeta_+\; e^{\II(\gamma_+-\gamma_1-\gamma_2)\zeta_+}\nonumber\\
&=& \delta_{x_1,x_2} \delta(\gamma_1-\gamma_2)\Theta(\gamma_1)|t|^{-2x_1}f_1(\mu)f_2(\gamma_1)
\exp\left(-2\gamma_1\ln(1+\mu r/t)/\mu\right)\nonumber\\
&=& \delta_{x_1,x_2} \delta(\gamma_1-\gamma_2)
\Theta(\gamma_1)f_1(\mu)f_2(\gamma_1)|t|^{-2x_1}(1+\mu r/t)^{-2\gamma_1/\mu}.
\label{corrpos}
\EEA
where in the third line two delta functions where recognised, and $f_1,f_2$ contain unspecified dependencies on $\mu$ and $\gamma_1$, 
respectively.\footnote{An eventual shift $\zeta_+ \mapsto \zeta_+ + c$, see (\ref{Ncovariance}), can be absorbed into the re-definition 
$\hat{\cal F}(\gamma_+)\,e^{-\gamma_+ c} \mapsto \hat{\cal F}(\gamma_+)$.} 
Analogously, for  $\lambda<0$, we have from (\ref{3.20},\ref{3.21}) 
\BEA 
F &=& \frac{|t|^{-2x}}{\pi\sqrt{2\pi}}\, {\hat G}_1(\mu)\int_{\R^2} \!\D \zeta_+ \D\zeta_-\; 
e^{-\II(\gamma_1+\gamma_2)\zeta_+}\, e^{-\II(\gamma_1-\gamma_2)\zeta_-}
\int_{\R}\!\D\gamma_+\,\Theta(\gamma_+){\hat {\cal F}}_-(\gamma_+)e^{+\gamma_+\lambda}e^{-\II\gamma_+\zeta_+}\nonumber\\
&=& \frac{|t|^{-2x}}{\pi\sqrt{2\pi}}\, {\hat G}_1(\mu)\int_{\R} \!\D\gamma_+\,\Theta(\gamma_+){\hat {\cal F}}_-(\gamma_+)e^{+\gamma_+\lambda}
\int_{\R} \!\D\zeta_{-}\; e^{-(\gamma_1-\gamma_2)\zeta_-}\int_{\R} \!\D\zeta_+\; e^{\II(-\gamma_+-\gamma_1-\gamma_2)\zeta_+}\nonumber\\
&=& \delta_{x_1,x_2} \delta(\gamma_1-\gamma_2)\Theta(-\gamma_1)|t|^{-2x_1}f_1(\mu)f_2(-\gamma_1)
\exp\left(-2\gamma_1\ln(1+\mu r/t)/\mu\right)\nonumber\\
&=& \delta_{x_1,x_2} \delta(\gamma_1-\gamma_2)
\Theta(-\gamma_1)f_1(\mu)f_2(-\gamma_1)|t|^{-2x_1}(1+\mu r/t)^{-2\gamma_1/\mu}.
\label{corrneg}
\EEA
Now, we discuss the meaning of the signs of $\lambda$, using (\ref{3.20}). We adopt the {\em convention that the mass $\mu >0$}. 
For $\lambda>0$, one must have that $\ln(1+\mu r/t)>0$, hence $r/t>0$. From the explicit expression (\ref{corrpos}), we also have $\gamma_1>0$.    
Conversely, for $\lambda<0$, one must have $\ln(1+\mu r/t)<0$, hence $r/t<0$ and from (\ref{corrneg}), we also have $\gamma_1<0$. 
Hence we have always $\gamma_1 r/t =\left| \gamma_1 r/t\right|>0$, independently of the sign of $\lambda$. 
Therefore, we can always write for the time-space argument 
\BD
\mu\frac{r}{t} = \frac{\mu}{\gamma_1} \frac{\gamma_1 r}{t} = \frac{\mu}{\gamma_1} \left|\frac{\gamma_1 r}{t}\right|
\ED
(if $\gamma_1\ne 0$) and we have identified the source of the non-analyticity in the two-point function, 
which rendered the na\"{\i}ve use of the meta-conformal Ward identities
(which would have led to (\ref{meta2})) inapplicable. Now, eqs.~(\ref{corrpos},\ref{corrneg}) combine as follows, which is or main result.

\noindent{\bf Proposition 2.} 
{\em With the convention that $\mu=\mu_1=\mu_2>0$, and if $\nu_1+\nu_2 >\frac{1}{2}$,
the two-point correlation function, co-variant under the representation (\ref{finitvarconf}), enhanced by (\ref{Ngenerator}), 
of the extended meta-conformal algebra 
$\wit{\mathfrak{conf}}(2)$, reads up to normalisation}
\BEQ \label{final}
\mathscr{C}_{12}(t,r)=\langle\phi_1(t,r)\phi_2(0,0)\rangle=
\delta_{x_1,x_2}\delta_{\gamma_1,\gamma_2}\,
|t|^{-2x_1}\left(1+\frac{\mu}{\gamma_1}\left|\frac{\gamma_1 r}{t}\right| \right)^{-2\gamma_1/\mu}.
\EEQ

This form has the correct symmetry $\mathscr{C}_{12}(t,r)=\mathscr{C}_{21}(-t,-r)$ under permutation of the scaling operators of a correlator. 
For $\gamma_1>0$ and $x_1>0$, the correlator decays to zero for $t\to\pm\infty$ or $r\to\pm\infty$.
This is analogous to ortho-conformal invariance (\ref{1C}), but distinct from a maximally asymmetric response function (\ref{2R}) as one finds for
a Schr\"odinger-invariant system.  

In the limit $\mu\to 0$, the extended meta-conformal algebra (\ref{commutators},\ref{Ncommutators}) 
contracts to the extended altern-Virasoro algebra $\wit{\mathfrak{altv}}(1)$, 
whose maximal finite-dimensional sub-algebra is the extended conformal Galilean algebra
$\wit{\mbox{\sc cga}}(1) = \mbox{\sc cga}(1)\oplus\mathbb{C}N$. We recover as a special limit case: 

\noindent{\bf Proposition 3.} \cite{Henkel15}
{\em If $\nu_1+\nu_2>\frac{1}{2}$, the two-point correlation function, co-variant under the extended conformal Galilean algebra 
$\wit{\mbox{\sc cga}}(1)$, reads up to normalisation}
\BEQ \label{final0}
\mathscr{C}_{12}(t,r)=\langle\phi_1(t,r)\phi_2(0,0)\rangle=
\delta_{x_1,x_2}\delta_{\gamma_1,\gamma_2}\,
|t|^{-2x_1} \exp\left( - \left|\frac{2\gamma_1 r}{t}\right| \right).
\EEQ

Summarising, we have shown that for time-space meta-conformal invariance, 
as well as for conformal galilean invariance, its $\mu\to 0$ limit, the co-variant
two-point correlators are given by eqs.~(\ref{final},\ref{final0}) and are explicitly {\em non-analytic} in the temporal-spatial variables. This
implies that any form of the Ward identities which implicitly assumes such an analyticity cannot be correct. In our construction of
physically sensible Ward identities, it was necessary to extend the Cartan sub-algebra to a higher rank. This points towards the possibility
for an extension of the meta-conformal and conformal galilean symmetries, in analogy to the embedding of the Schr\"odinger algebra
$\mathfrak{sch}(1)\subset \mathfrak{conf}(3)$ into a conformal algebra in three dimensions \cite{Henkel03a}. 
These extensions and their physical consequences remain to be found. 

{\small \noindent {\bf Acknowledgement:} This work was supported by PHC Rila and from the Bulgarian National Science Fund Grant DFNI - T02/6.}

\end{document}